\documentclass[aps,pra,superscriptaddress,amsmath,amssymb,showpacs,twocolumn,notitlepage]{revtex4-1}
\usepackage{amssymb}
\usepackage{amsmath}
\usepackage{graphicx}
\usepackage{bm}
\usepackage{bbm}
\usepackage{color}
\usepackage{times}
\usepackage{xcolor}

\renewcommand{\eqref}[1]{(\ref{#1})}

\graphicspath{{./Figures/}}

\DeclareMathOperator{\Ima}{Im}

\begin{document}

\title{Graphene-based amplification and tuning of near-field radiative\\heat transfer between dissimilar polar materials}

\author{Riccardo Messina}
\email{riccardo.messina@umontpellier.fr}
\affiliation{Laboratoire Charles Coulomb (L2C), UMR 5221 CNRS-Universit\'{e} de Montpellier, F-34095 Montpellier, France}
\author{Philippe Ben-Abdallah}
\email{pba@institutoptique.fr}
\affiliation{Laboratoire Charles Fabry, UMR 8501, Institut d'Optique, CNRS, Universit\'{e} Paris-Saclay, 2 Avenue Augustin Fresnel, 91127 Palaiseau Cedex, France}
\affiliation{Universit\'{e} de Sherbrooke, Department of Mechanical Engineering, Sherbrooke, PQ J1K 2R1, Canada.}
\author{Brahim Guizal}
\email{brahim.guizal@umontpellier.fr}
\affiliation{Laboratoire Charles Coulomb (L2C), UMR 5221 CNRS-Universit\'{e} de Montpellier, F- 34095 Montpellier, France}
\author{Mauro Antezza}
\email{mauro.antezza@umontpellier.fr}
\affiliation{Laboratoire Charles Coulomb (L2C), UMR 5221 CNRS-Universit\'{e} de Montpellier, F- 34095 Montpellier, France}
\affiliation{Institut Universitaire de France, 1 rue Descartes, F-75231 Paris Cedex 05, France}

\begin{abstract}
The radiative heat transfer between two dielectrics can be strongly enhanced in the near field in the presence of surface phonon-polariton resonances. Nevertheless, the spectral mismatch between the surface modes supported by two dissimilar materials is responsible for a dramatic reduction of the radiative heat flux they exchange. In the present paper we study how the presence of a graphene sheet, deposited on the material supporting the surface wave of lowest frequency, allows to widely tune the radiative heat transfer, producing an amplification factor going up to one order of magnitude. By analyzing the Landauer energy transmission coefficients we demonstrate that this amplification results from the interplay between the delocalized plasmon supported by graphene and the surface polaritons of the two dielectrics. We finally show that the effect we highlight is robust with respect to the frequency mismatch, paving the way to an active tuning and amplification of near-field radiative heat transfer in different configurations.
\end{abstract}

\maketitle

\section{Introduction}

Improving radiative heat exchanges between two bodies separated by a gap is a longstanting problem in physics. At large separation distance (i.e. far-field regime) energy exchanges result exclusively from propagative photons emitted by these media and the blackbody limit~\cite{Planck} sets the maximum heat flux which can be exchanged between two objects. However, at subwavelength distances (i.e. near-field regime) the situation radically changes~\cite{Rytov,PoldervH}. Indeed, at this scale, the evanescent photons which remain confined near the surface of materials~\cite{Eckardt} are the main contributors to the heat transfer~\cite{JoulainSurfSciRep05,Volokitin1} by tunneling through the separation gap. It results from this transport a significant heat flux increase~\cite{HargreavesPLA69,KittelPRL05,HuApplPhysLett08,NarayanaswamyPRB08,RousseauNaturePhoton09,ShenNanoLetters09,KralikRevSciInstrum11,
OttensPRL11,vanZwolPRL12a,vanZwolPRL12b,KralikPRL12,KimNature15,StGelaisNatureNano16,SongNatureNano15,KloppstecharXiv,WatjenAPL16}. In the presence of resonant surface modes~\cite{JoulainSurfSciRep05}, a continuum of hyperbolic modes~\cite{BenAbdallahPRL12} or surface Bloch waves~\cite{BenAbdallahAPL10}, the radiative heat exchanges can drastically surpass by several orders of magnitude the prediction of Planck's blackbody theory. However, when the two media in interaction are dissimilar, the spectral mismatch between their optical properties limits dramatically the amount of energy they can exchange between each other~\cite{BenAbdallahPRB10}. To limit this effect, composite systems made with a single or several graphene sheets have been suggested~\cite{Persson,Volokitin,Svetovoy1,Ilic1,Ilic2,Messina1,Messina2,Lim2,Phan,Liu3,Liu1,Svetovoy2,Drosdoff, Zhang1,Lim1,Chang,Song,Yin,Zheng,Zhao1,Simchi,Lim3,Zhao2,Shi}. These systems exploit the exceptional optical properties of graphene~\cite{Geim1,Geim2}. More specifically, the radiative heat transfer between suspended graphene sheets has been analyzed~\cite{Ilic1,Yin}, as well as in configurations where graphene is deposited either on dielectric substrates~
\cite{Persson,Volokitin,Svetovoy1,Messina1,Lim2,Liu3,Zhao1,Shi} or on metamaterials~\cite{Liu1,Drosdoff,Zhang1,Chang,Song,Zhao2}. Beside these fundamental develpements, the graphene sheets has also been considered for several applicative purposes, such as thermophotovoltaic conversion~\cite{Ilic2,Messina2,Svetovoy2,Lim1,Lim3}, thermal rectification~\cite{Zheng} and heat transfer amplification~\cite{Simchi}.

In this paper we investigate the role that a graphene sheet can play on the near-field heat exchanges between two planar media which support two surface waves at two different frequencies in the Planck window where the near-field exchanges take place. To this aim we consider two polar materials and we analyze, using the Landauer-like theory of radiative heat exchanges~\cite{BenAbdallahPRB10}, the net heat flux exchanged between these media when a graphene is deposited on one of polar materials. The paper is structured as follows. In Sec.~\ref{SecPhys}, we present our physical system, introduce the optical properties of involved materials and remind the definition heat flux exchanged both in near field and in far field between two planar media. In Sec.~\ref{SecFlux} we calculate this flux with respect to the separation distances betweeen the two polar media and with respect to the chemical potential of graphene. To quantify the role played by the graphene sheet we also introduce an amplification coefficient of heat flux due to the presence of graphene and we show that under appropriate conditions the latter can significantly amplify energy exchanges between the two polar materials despite of their spectral mismatch. We also demonstrate that the heat flux can be controlled by tuning the chemical potential of graphene. Next, to get insight on the physical origin of the amplification we analyze in Sec.~\ref{SecTransmission} in the frequency-wavevector plane the Landauer transmission coefficients, that is the coupling efficiency of modes supported by the two media, in the presence or not of graphene with respect to the separation distance and the doping level. We then show in Sec.~\ref{SecRobustness} that the amplification we describe is stable with respect to the frequency mismatch between the two dielectric substrates. We finally summarize our results in Sec.~\ref{SecConclusions}.

\section{Physical system}\label{SecPhys}

\begin{figure}[t!]
\includegraphics[width=0.48\textwidth]{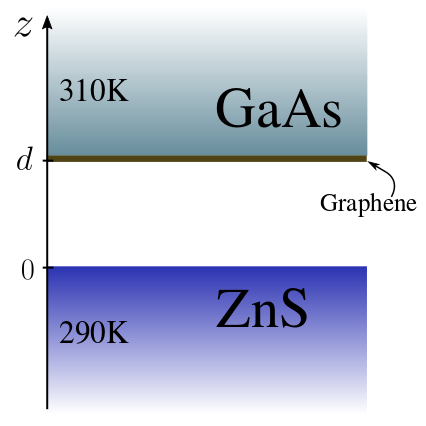} 
\caption{Geometry of the system. Two planar slabs of infinite thickness, made of ZnS and GaAs respectively, are separated by a distance $d$. A graphene sheet is deposited at $z=d$, i.e. at the vacuum-GaAs interface. The temperatures of the two slabs are fixed at $T_1=290\,$K and $T_2=310\,$K throughout the paper.}
\label{Sys}
\end{figure}

The physical system we consider, represented in Fig.~\ref{Sys}, is made of two parallel planar slabs of infinite thickness. We assume that slab 1, made of zinc sulfide (ZnS), is kept at a temperature $T_1=290\,$K, while slab 2, made of gallium arsenide (GaAs), is kept at $T_2=310\,$K, so that the Planck window is centered around $\lambda=10\,\mu$m ($\omega\simeq1.8\times10^{14}\,$rad/s). For the distance between the two slabs, noted with $d$, we will consider the region going from 10\,nm to $3\,\mu$m, in order to explore the transition between near and far field. Starting from this reference configuration, we will first study the effect of one single layer of graphene deposited on the vacuum\,-\,GaAs interface, as shown in Fig.~\ref{Sys}.

Let us now focus on the optical description of the materials involved in the problem. We describe both GaAs and ZnS by means of a Drude-Lorentz model
\begin{equation}
 \varepsilon(\omega) = \varepsilon_\text{inf}\, \frac{\omega^2 - \omega_\text{L}^2 + i \gamma \omega}{\omega^2 - \omega_\text{T}^2 + i \gamma \omega},
\label{Drude}\end{equation}
the model parameters being~\cite{Palik} $\varepsilon_\text{inf} = 5.7$, $\omega_\text{L} = 0.66 \times 10^{14}\,$\text{rad/s}, $\omega_\text{T} = 0.53 \times 10^{14}\,$\text{rad/s} and $\gamma = 1.28 \times 10^{12}\,$\text{rad/s} for ZnS, whereas $\varepsilon_\text{inf} = 11.0$, $\omega_\text{L} = 0.55 \times 10^{14}\,$\text{rad/s}, $\omega_\text{T} = 0.51 \times 10^{14}\,$\text{rad/s} and $\gamma = 4.52 \times 10^{11}\,$\text{rad/s} for GaAs. This model predicts for both materials the existence of a surface phonon-polariton resonance, having frequency $\omega_1 \simeq 0.65 \times 10^{14}\,$\text{rad/s} for ZnS, and a lower frequency $\omega_2 \simeq 0.55 \times 10^{14}\,$\text{rad/s} for GaAs.

The optical properties of graphene will be described in terms of a 2D conductivity $\sigma(\omega)$. Following Ref.~\onlinecite{FalkovskyJPhysConfSer08}, this can be written as a sum of an intraband (Drude) and an interband contributions, respectively given by 
\begin{eqnarray}\label{Sigma}&&\sigma_D(\omega)=\frac{i}{\omega+\frac{i}{\tau}}\frac{2e^2k_BT}{\pi\hbar^2}\log\Bigl(2\cosh\frac{\mu}{2k_BT}\Bigr),\\
&&\nonumber\sigma_I(\omega)=\frac{e^2}{4\hbar}\Bigl[G\Bigl(\frac{\hbar\omega}{2}\Bigr)+i\frac{4\hbar\omega}{\pi}\int_0^{+\infty}\frac{G(\xi)-G\bigl(\frac{\hbar\omega}{2}\bigr)}{(\hbar\omega)^2-4\xi^2}\,d\xi\Bigr],\label{sigma}\end{eqnarray}
where $G(x)=\sinh(x/k_BT)/[\cosh(\mu/k_BT)+\cosh(x/k_BT)]$. The conductivity depends explicitly on the temperature $T$ of the graphene sheet, for which we have chosen the same value $T=310\,$K of the GaAs substrate. Besides, Eq.~\eqref{sigma} contains the relaxation time $\tau$, which we have fixed (following Ref.~\onlinecite{JablanPRB09}) to the value $\tau=10^{-13}\,$s. Finally, the conductivity depends on the chemical potential $\mu$, allowing to actively tune the optical response of graphene and, in turn, the radiative heat transfer between the two structures.

We now need the explicit expression of the radiative heat transfer per unit area exchanged between the two structures. It is convenient to express this under the form of a Landauer decomposition~\cite{BenAbdallahPRB10}
\begin{equation}
\varphi = \int_0^\infty\frac{d\omega}{2\pi} \hbar\omega\, n_{21}(\omega) \sum_p\int \frac{d^2 \mathbf{k}}{(2\pi)^2} \mathcal{T}_p(\omega,\mathbf{k}),
\label{flux}
\end{equation}
where $n_{\alpha\beta}(\omega)=n_\alpha(\omega)-n_\beta(\omega)$ is the difference between the two mean photon occupation numbers $n_\alpha(\omega)=(\exp[\hbar\omega /k_B T_\alpha]-1)^{-1}$, with $\alpha=1,2$. The decomposition in Eq.~\eqref{flux} describes the radiative heat flux as a sum of contribution coming from each field mode, identified by the frequency $\omega$, the parallel wavevector $\mathbf{k}=(k_x,k_y)$ and the polarization $p$, which can be TE (transverse electric) or TM (transverse magnetic): each mode transports an energy $\hbar\omega$, multiplied by a transmission coefficient $\mathcal{T}_p(\omega,\mathbf{k})$, taking values between 0 and 1. In the case of two parallel planar slabs, this quantity reads (the dependence on frequency and wavevector is implicit)
\begin{equation}
\begin{split} 
\mathcal{T}_p &= \begin{cases} {\displaystyle \frac{(1-\left|\rho_{1,p}\right|^2)(1-\left|\rho_{2,p}\right|^2) }{ \left|1-\rho_{1,p}\,\rho_{2,p}e^{2ik_zd}\right|^2}}, & k < \frac{\omega}{c},\\
\\
{\displaystyle \frac{4\Ima\left(\rho_{1,p}\right)\Ima\left(\rho_{2,p}\right)e^{-2\Ima(k_z)d}}{\left|1-\rho_{1,p}\,\rho_{2,p}e^{-2\Ima(k_z)d}\right|^2}}, & k > \frac{\omega}{c}, \end{cases}\\
\label{Twk}
\end{split}
\end{equation}
where $k_z=\sqrt{\omega^2/c^2-\mathbf{k}^2}$ is the normal component of the wavevector in vacuum, while $\rho_{i,p}$ is the reflection coefficient of body $i=1,2$ for polarization $p$, taking into account, in the case of body 2, the presence of graphene. We stress that the reflection coefficient $\rho_{2,p}$ contains the graphene conductivity, and thus depends on the temperature $T_2$ and on the chemical potential $\mu$.

\section{Amplification and tuning of\\radiative heat transfer}\label{SecFlux}

\begin{figure*}[t!]
\includegraphics[width=\textwidth]{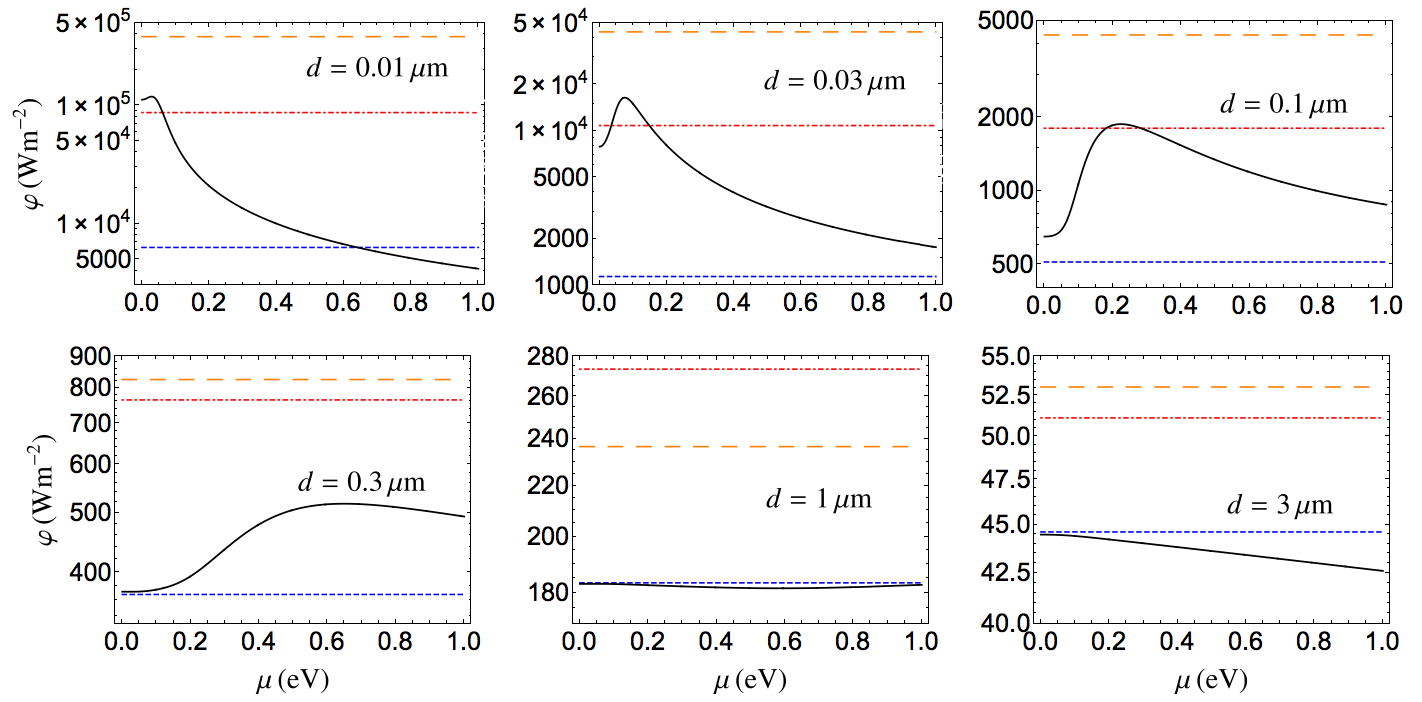} 
\caption{Radiative heat flux $\varphi$ per unit area as a function of the graphene chemical potential $\mu$ for six different values of the distance $d$ (as indicated in each panel). The three horizontal lines in each curve correspond to three reference values in the absence of graphene: ZnS--ZnS (orange long-dashed line), GaAs--GaAs (red dot-dashed line) and ZnS--GaAs (blue dashed line).}
\label{Grid}
\end{figure*}

We now have all the ingredients needed to analyze the heat transfer given by Eq.~\eqref{flux} as a function of both the distance $d$ and the graphene chemical potential $\mu$. As far as the former is concerned, we are going to explore the region $d\in[10\,\text{nm},3\,\mu\text{m}]$, fully catching the near-field behavior and the transition toward the far field. Concerning the chemical potential, we are going to restrict our analysis to the range $\mu\in[0,1]\,\text{eV}$, containing physically accessible values.

We show our first set of results in Fig.~\ref{Grid}, where the flux $\varphi$ is plotted as a function of the chemical potential $\mu$ for six different values of the distance $d$. In each panel, the $\mu$-depending flux is compared with three reference values in the absence of graphene, i.e. the two configurations GaAs--GaAs and ZnS--ZnS of equal dielectrics as well as the scenario ZnS--GaAs, obtained just removing the graphene sheet.

A first glimpse of the six curves already gives an idea of the possibilities offered by the presence of graphene in terms of manipulation of the radiative flux. In particular, we observe not only that the flux can be monotonic or not with respect to $\mu$ depending on the distance considered, but also that the chemical potentials maximizing or minimizing the transfer are as well functions of $d$. As a general feature, we note however that (at least in the window of $\mu$ taken into account) approaching the far field reduces the degree of variation of the flux with respect to $\mu$.

\begin{figure}[b!]
\includegraphics[width=0.48\textwidth]{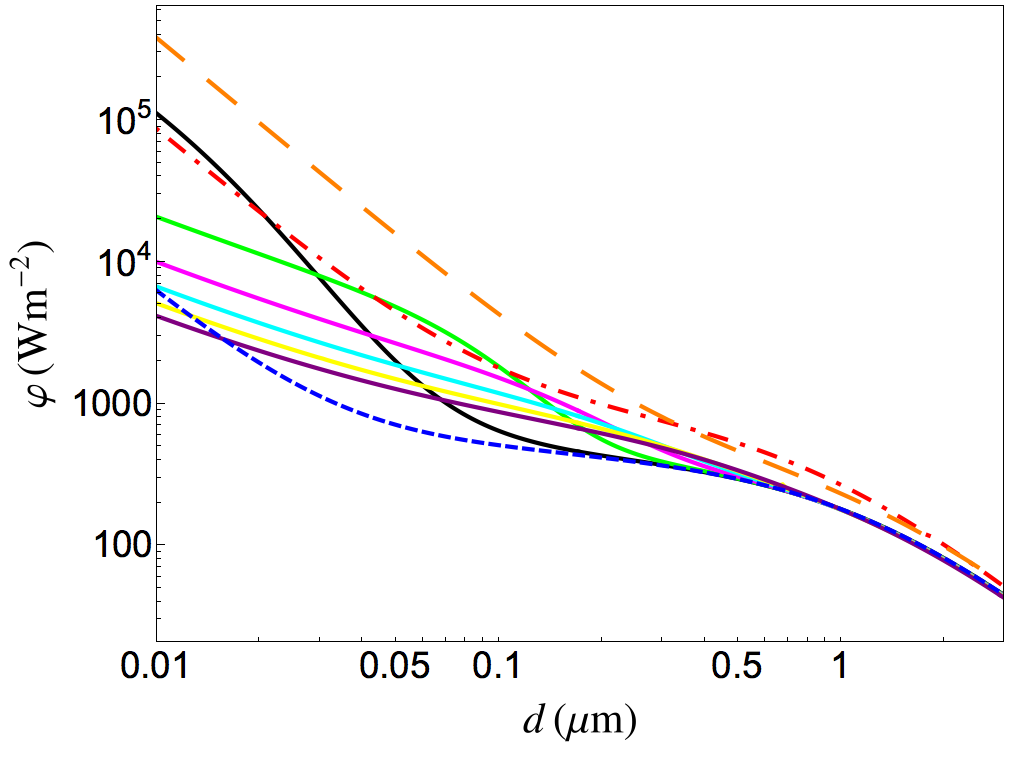} 
\caption{Radiative flux $\varphi$ per unit area as a function of the distance $d$ for six different values of the graphene chemical potential $\mu$. From top to bottom at $d=10\,$nm, the curves correspond to $\mu=0\,$eV (black), 0.2\,eV (green), 0.4\,eV (magenta), 0.6\,eV (cyan), 0.8\,eV (yellow) and 1.0\,eV (purple). We also show three lines corresponding to reference values in the absence of graphene: ZnS--ZnS (orange long-dashed line), GaAs--GaAs (red dot-dashed line) and ZnS--GaAs (blue dashed line).}
\label{Diffmu}
\end{figure}

It is also instructive to consider the three reference values. We first remark that, not surprisingly, the values corresponding to couples of equal dielectrics always give a flux much higher than the configuration ZnS--GaAs, characterized by a surface-resonance frequency mismatch. More interestingly, for some values of the distance, tuning the chemical potential allows to go beyond the value of the flux corresponding to two GaAs substrates. This proves that the presence of graphene is not only able to permit a large variation and amplification of the flux through its chemical potential, but also to fully compensate the mismatch between the resonance frequencies of the two dielectrics.

This feature is more manifest in the complementary plot given in Fig.~\ref{Diffmu}, where the flux is shown as a function of the distance $d$ for five different values of the chemical potential. We first confirm that the largest possible tuning (and amplification) is realized at the smallest distance, while all the curves corresponding to different values of $\mu$ converge to each other and to the configuration corresponding to the absence of graphene (blue dashed curve) when moving to the far field. Moreover, coherently with what was observed before, the different solid lines cross each other, showing that for each $d$ the highest and lowest fluxes are realized for different chemical potentials. Finally, we clearly highlight two regions of distances where even the flux between two GaAs substrates is surpassed.

\begin{figure}[t!]
\includegraphics[width=0.48\textwidth]{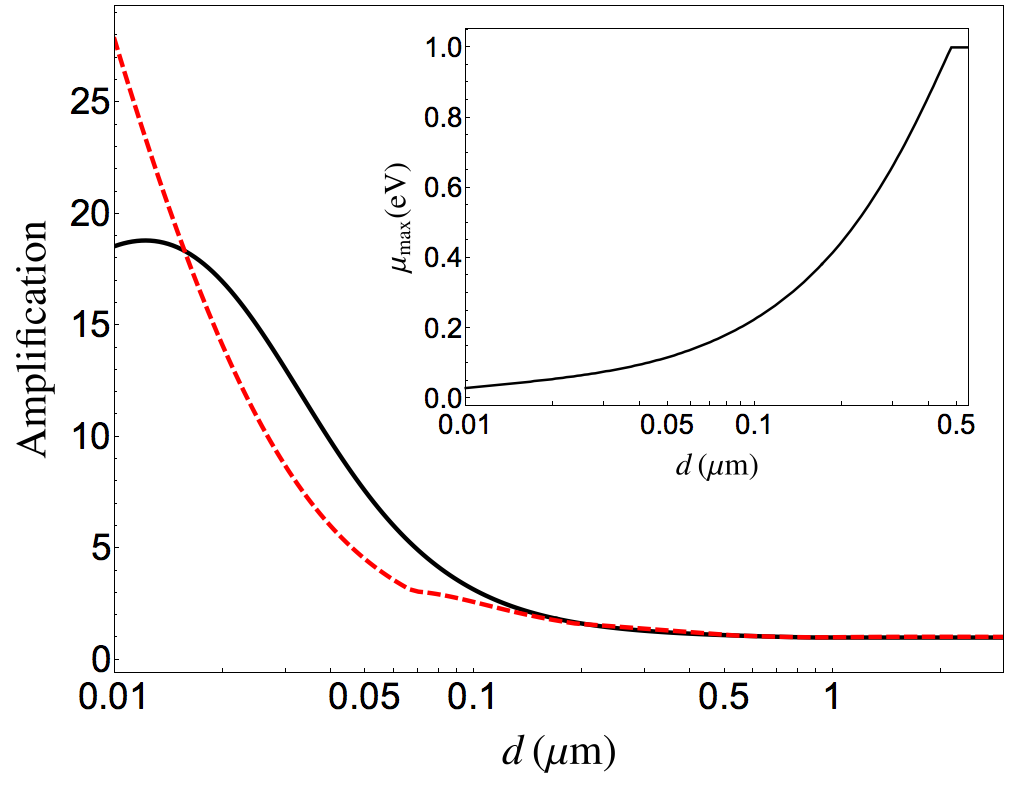} 
\caption{Radiative-heat-flux amplification factor with respect to the graphene chemical potential $\mu$ as a function of the distance $d$. The red dashed curve corresponds to the ratio between the highest and lowest values of the flux in the presence of graphene, while the black curve corresponds to the ratio between the highest value of $\varphi$ in the presence of graphene and the flux in the ZnS--GaAs configuration without graphene. In the inset the chemical potential $\mu_\text{max}$ realizing the maximum flux in the presence of graphene is shown as a function of the distance $d$.}
\label{Amplification}
\end{figure}

Inspired by the results presented so far, it is interesting to give an overall image of the possibilities offered by the presence of graphene in terms of tuning and amplification of the flux. With this respect, two complementary views are possible. On one hand, the ratio between the maximum and minimum values of $\varphi$ (with respect to the chemical potential $\mu$) can be plotted versus the distance $d$: this quantity tells us how much we can tune the flux by externally acting on the chemical potential. This ratio corresponds to the red dashed line in Fig.~\ref{Amplification}. On the other hand, one can calculate the ratio between the maximum value of the flux with respect to $\mu$ and the reference value in the ZnS--GaAs configuration, i.e. in the absence of graphene. This second ratio describes how the presence of graphene is able to amplify the standard radiative flux between two dissimilar dielectrics by compensating the mismatch between the two different resonance frequencies. This quantity corresponds to the black line in Fig.~\ref{Amplification}. The analysis of these two curves shows that in both cases we have a remarkable amplification factor which can go beyond one order of magnitude. More specifically, Fig.~\ref{Amplification} clearly shows that the effect we highlight is a near-field effect. As a matter of fact, starting from $d=500\,$nm, i.e. when moving toward the far-field region, the two curves join each other and tend to 1, which means to an almost flat value of $\varphi$ as a function of $\mu$, in agreement with the last panels of Fig.~\ref{Grid}. The inset of Fig.~\ref{Amplification} shows the value $\mu_\text{max}$ of the chemical potential realizing the maximum value of the flux. This quantity is plotted for $d\lesssim500\,$nm, i.e. in the region of distances showing a significant amplification. The curve shows that the value $\mu_\text{max}$ saturates for $d\simeq500\,$nm at the maximum value of 1\,eV imposed in our calculation. One must therefore bear in mind that also the two amplification curves shown in the main part of Fig.~\ref{Amplification} are influenced by this choice. We will comment further on this point in the discussion of the transmission coefficients given below.

\section{Transmission coefficient and\\spectral flux}\label{SecTransmission}

\begin{figure*}[t!]
\includegraphics[height=0.29\textwidth]{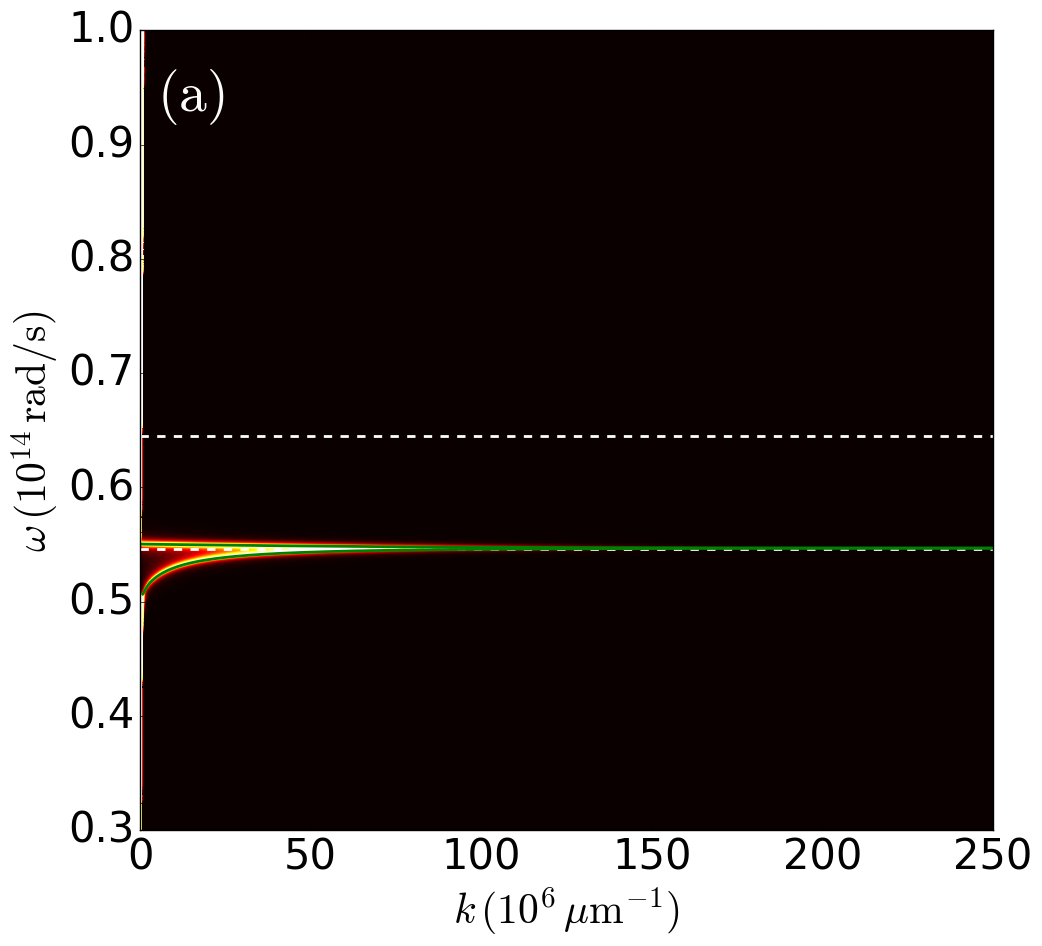}\quad\includegraphics[height=0.29\textwidth]{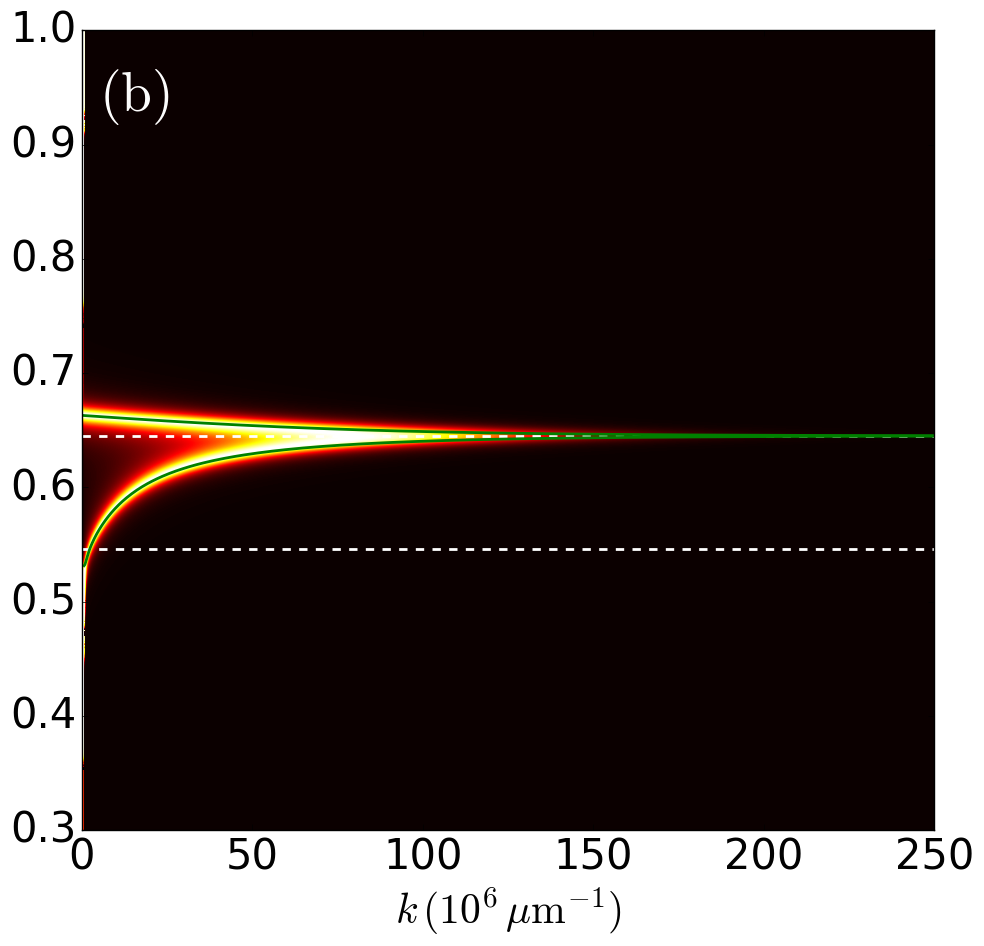}\quad\includegraphics[height=0.29\textwidth]{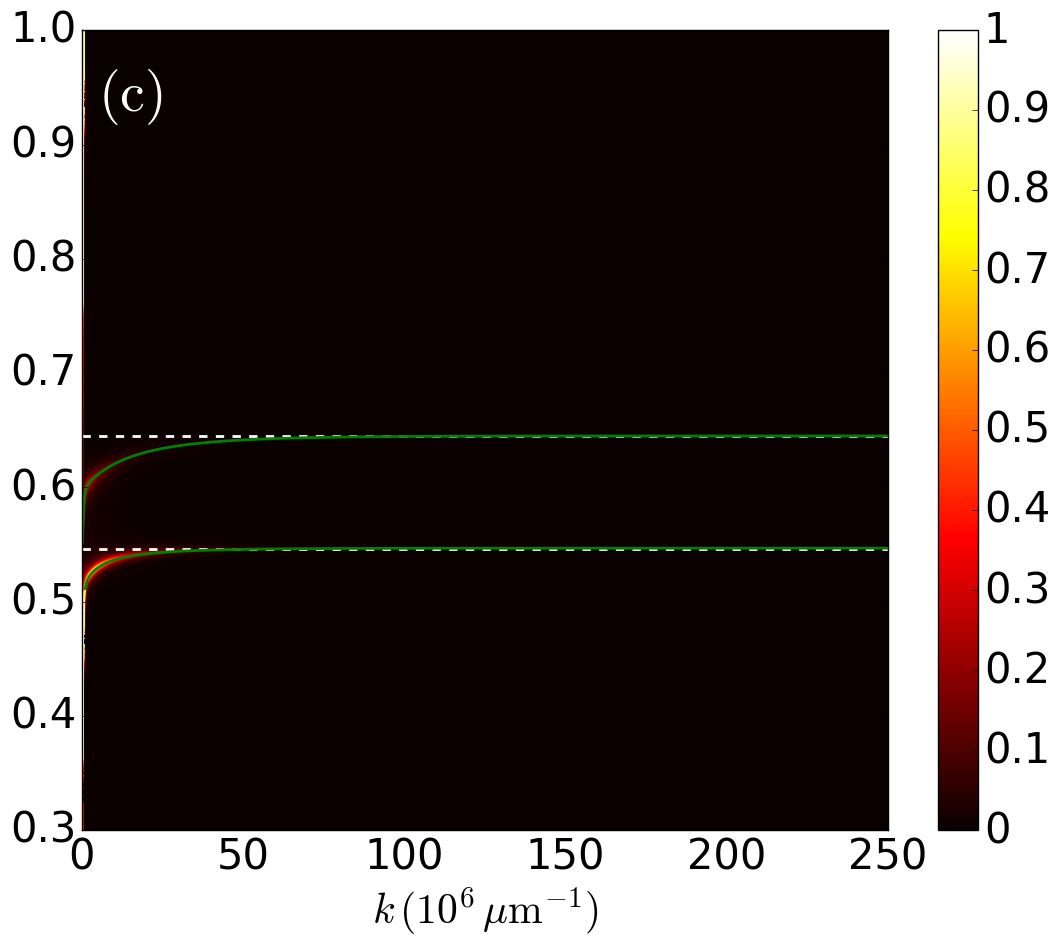}\\
\caption{Landauer transmission coefficient $\mathcal{T}_p(\omega,\mathbf{k})$ in the $(k,\omega)$ plane for the three reference dielectric--dielectric configurations in the absence of graphene for $d=20\,$nm in TM polarization: (a) GaAs--GaAs, (b) ZnS--ZnS and (c) ZnS--GaAs. The horizontal lines correspond to the resonance frequencies of GaAs and ZnS. The green lines describe the dispersion relation of the cavity surface modes.}
\label{Landauer}
\end{figure*}

To get more insight into the physics behind this tuning and amplification of radiative heat transfer we now focus on the analysis of the Landauer transmission coefficient $\mathcal{T}_p(\omega,\mathbf{k})$ for several configurations both in the absence and in the presence of graphene. As stated above, this quantity, always between 0 and 1, describes the rate of participation of the mode having polarization $p$, frequency $\omega$ and wavevector $\mathbf{k}$ to the energy exchange. In the following we focus only on TM polarization, since it is well known that this is the one mainly contributing to the amplification of radiative heat transfer in the near field~\cite{JoulainSurfSciRep05}. To start with, we focus on the distance $d=20\,$nm, well within the near-field region, and we show in Fig.~\ref{Landauer} the Landauer coefficients associated with the three standard dielectric--dielectric configurations, namely GaAs--GaAs, ZnS--ZnS and ZnS--GaAs. Figures \ref{Landauer}(a) and \ref{Landauer}(b) show a scenario typical in the literature of near-field radiative heat transfer. We see two branches (symmetric and antisymmetric) of surface modes, converging to a horizontal asymptote corresponding to the frequencies of the surface resonances of the two materials. We observe that the branches associated with GaAs are thinner and are limited to smaller values of $k$. This stems from the fact that GaAs has smaller losses than ZnS, as manifest from the parameters given after Eq.~\eqref{Drude}. Figure \ref{Landauer}(c) shows the transmission coefficient for the ZnS--GaAs configuration. We immediately see that, although the resonance frequencies are relatively close to each other, the mismatch produces a remarkable decoupling, reducing dramatically the number of modes effectively participating to the exchange, and thus the total integrated flux.

\begin{figure*}[t!]
\includegraphics[height=0.37\textwidth]{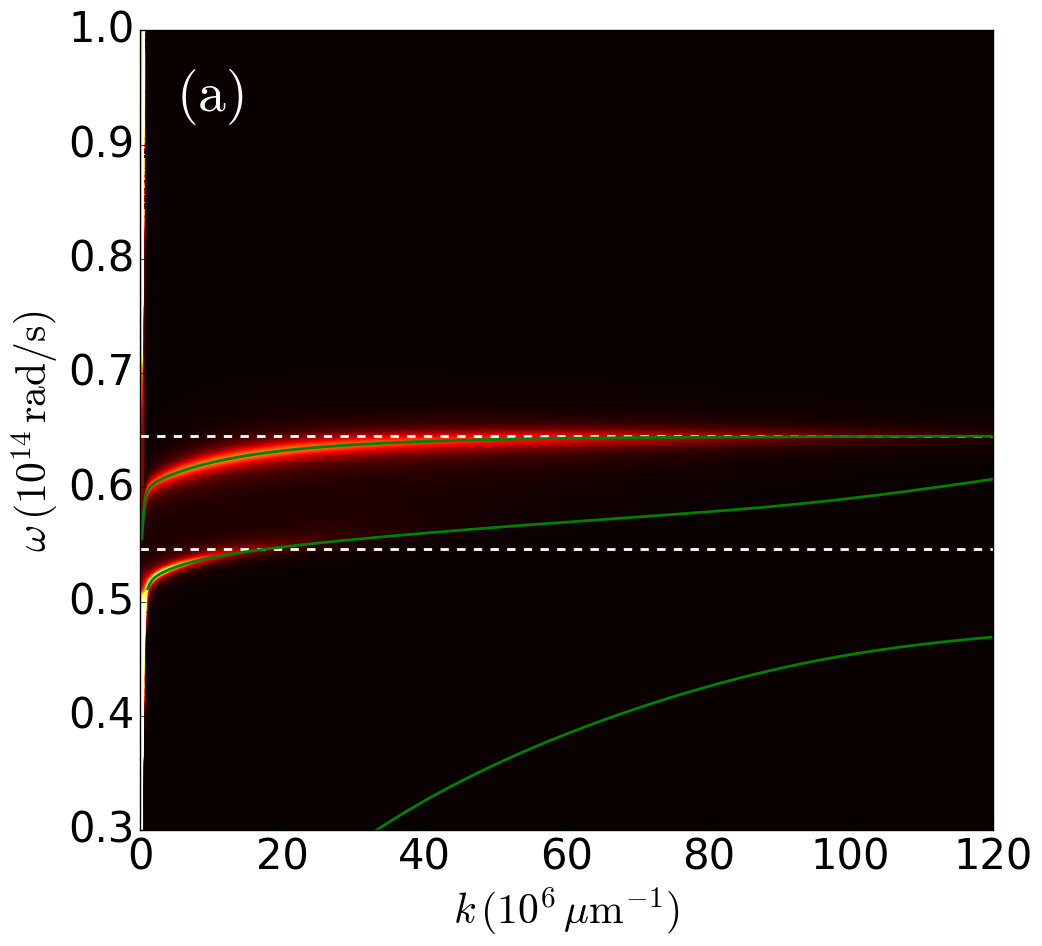}\hspace{1.5cm}\includegraphics[height=0.37\textwidth]{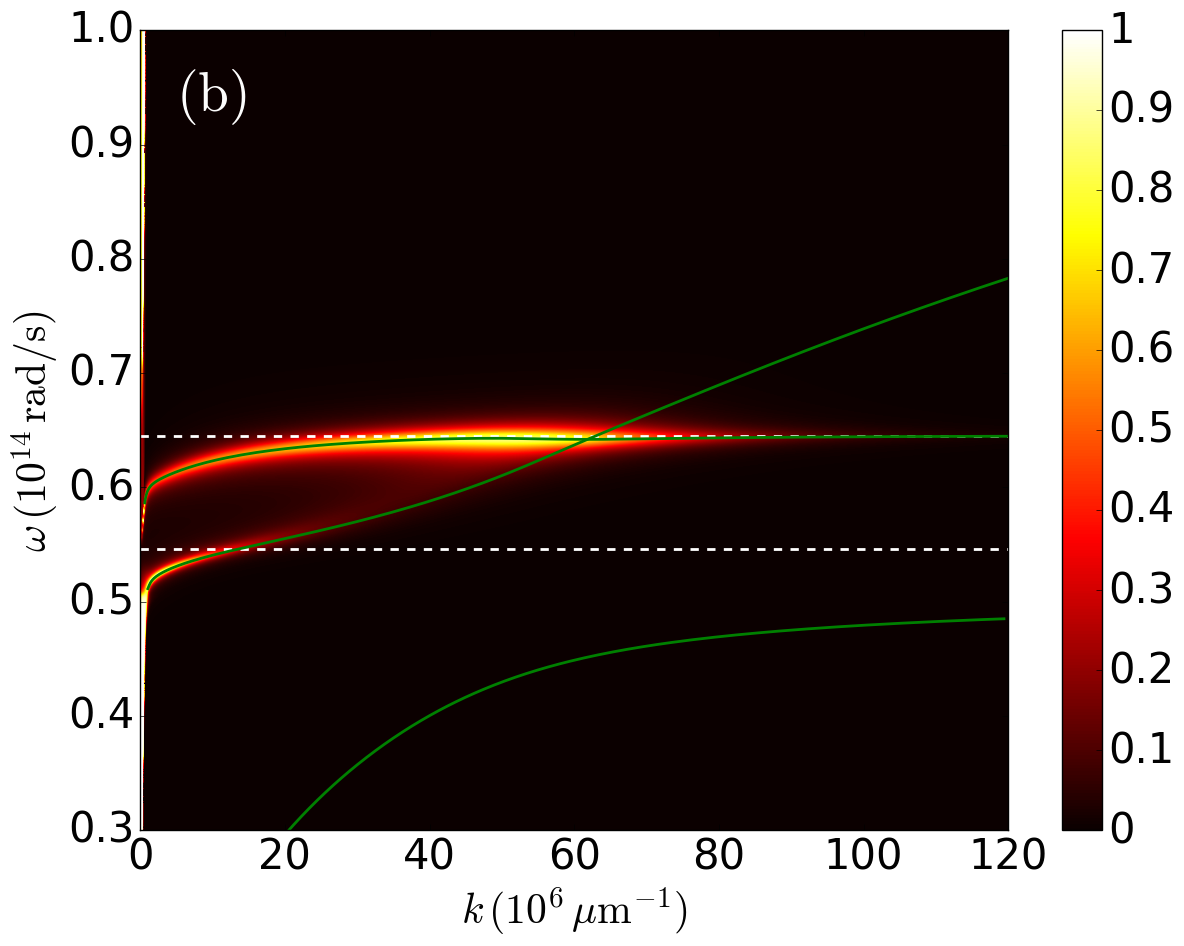}\\
\hspace{-0.8cm}\includegraphics[height=0.37\textwidth]{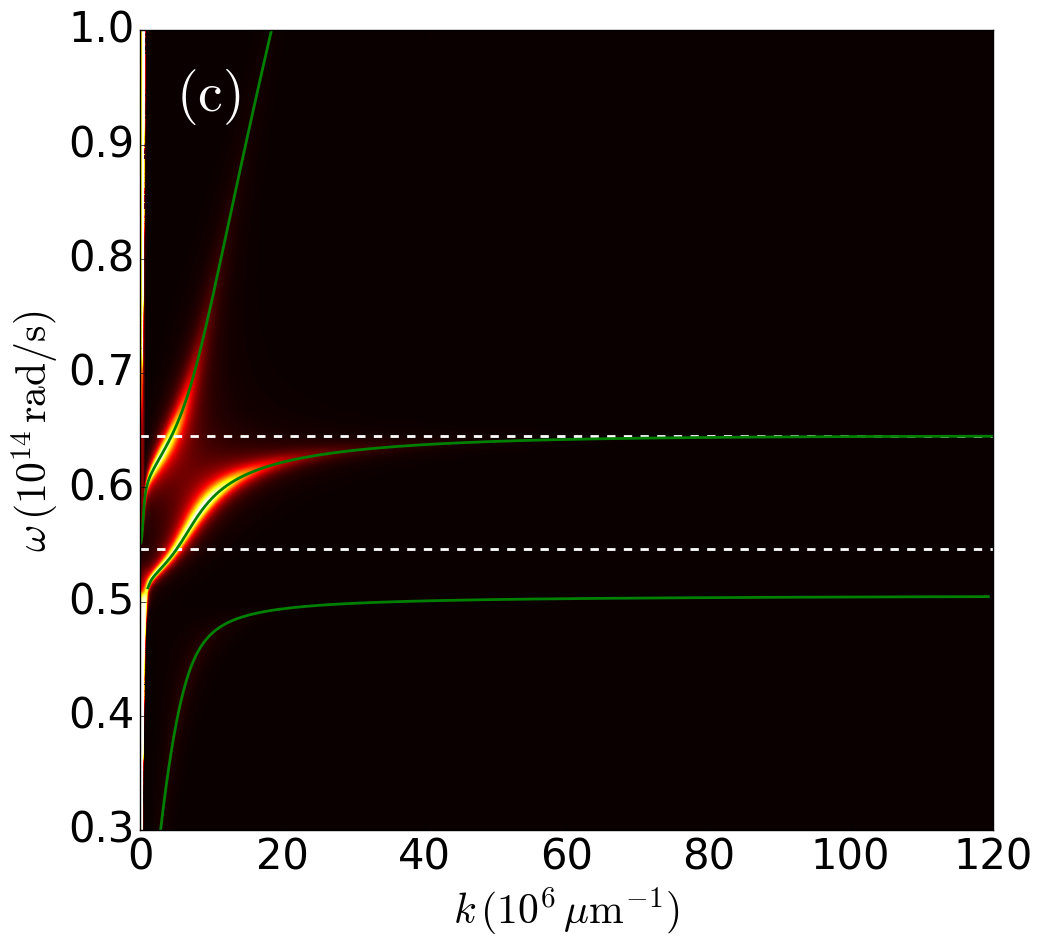}\hspace{1.5cm}\includegraphics[height=0.37\textwidth]{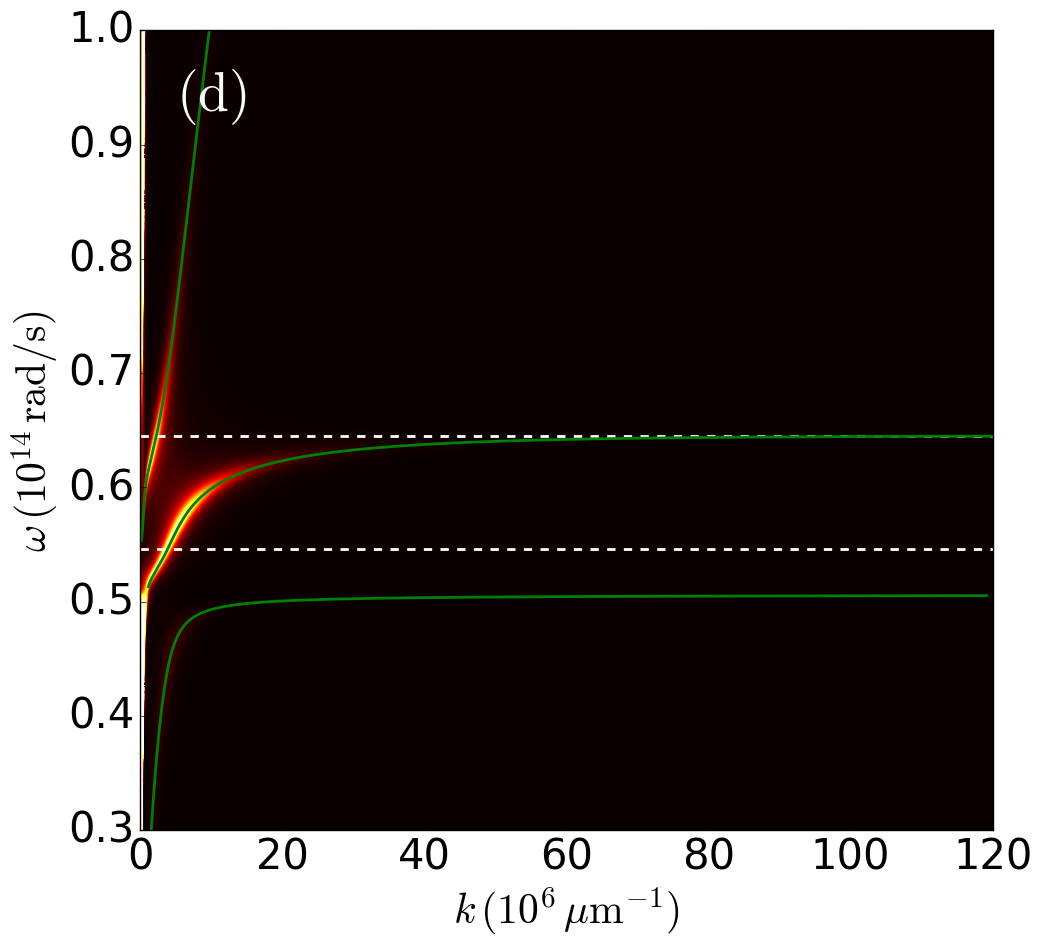}\\
\caption{Landauer transmission coefficient $\mathcal{T}_p(\omega,\mathbf{k})$ in the $(k,\omega)$ plane in the presence of graphene for $d=20\,$nm in TM polarization. The four panels correspond to different values of the chemical potential: (a) $\mu=0\,$eV, (b) $\mu\simeq\mu_\text{max}=0.05\,$eV, (c) $\mu=0.5\,$eV and (d) $\mu=1\,$eV. The horizontal lines correspond to the resonance frequencies of GaAs and ZnS. The green lines describe the dispersion relation of the cavity surface modes.}
\label{Landauer_g}
\end{figure*}

\begin{figure*}[t!]
\includegraphics[width=\textwidth]{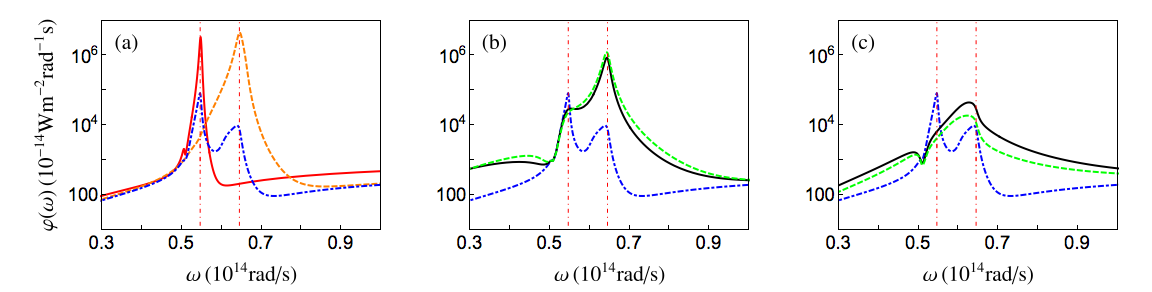} 
\caption{Spectral flux $\varphi(\omega)$ for several different configurations. Panel (a) shows three standard dielectric--dielectric configurations: GaAs--GaAs (red curve), ZnS--ZnS (orange dashed curve) and ZnS--GaAs (blue dot-dashed line). This last curve is compared in panels (b) and (c) to spectral fluxes in the presence of graphene. In panel (c) we have $\mu=0$ (black solid line) and $\mu=0.05\,$eV (green dashed line), while in panel (d) $\mu=0.5\,$eV (black solid line) and $\mu=1\,$eV (green dashed line) are shown. In the three panels the vertical red dot-dashed lines correspond to the two resonances $\omega_1$ and $\omega_2$ of ZnS and GaAs, respectively.}
\label{Grid_20nm}
\end{figure*}

We now turn our attention to the transmission coefficients in the presence of graphene, by considering in Fig.~\ref{Landauer_g} four different values of the chemical potential, namely $\mu=0\,$eV, 0.05\,eV (close to the value $\mu_\text{max}$ realizing the highest flux for $d=20\,$nm), 0.5\,eV and 1\,eV. As a general remark, by comparing these plots with Fig.~\ref{Landauer}(c), we observe that all the considered values of $\mu$ clearly increase the number of modes contributing to the flux. Besides, we see that for $\mu=0.05\,\text{eV}\simeq\mu_\text{max}$ the presence of graphene creates a region of modes with $\omega\simeq\omega_1$ (the resonance frequency of ZnS) and relatively high wavevector having a non-negligible value of $\mathcal{T}_p(\omega,\mathbf{k})$. In fact, it is not only important to observe the increased number of modes, but also their typical wavevector, since each mode participates to the heat transfer between two planar slabs through an additional factor $k$, namely from the Jacobian when moving to polar coordinates in the $(k_x,k_y)$ plane. For higher values of $\mu$ [Figs.~\ref{Landauer_g}(b) and \ref{Landauer_g}(c)] the branches of resonant modes manifestly move toward smaller values of $k$, reducing the total effect.

This transition as a function of $\mu$ and in particular the existence of an optimum chemical potential $\mu_\text{max}$ can be explained on the basis of the optical properties of graphene. It is well known that a suspended sheet of graphene has a delocalized surface resonance mode in TM polarization whose dispersion relation does not have a horizontal asymptote (as in the case of phonon-polaritons for dielectrics and plasmons for metals), but behaves as $\sqrt{k}$ for small wavevectors. As discussed for instance in Ref.~\onlinecite{Messina1}, when a graphene sheet is deposited on a dielectric substrate supporting a phonon-polariton resonance, we have a strong coupling between the two surface resonances, producing two non-crossing branches. The one at higher frequencies inherits the $\sqrt{k}$ behavior typical of graphene. We see a trace of this in the transmission coefficient shown in Fig.~\ref{Landauer_g}(b) corresponding to optimum heat transfer. Starting with a simplified analysis, we can say that the lower branch, associated to GaAs alone, which would have without graphene a horizontal asymptote at the GaAs resonance frequency, is shifted thanks to the presence of graphene toward higher frequencies and, thanks to the positive $d\omega/dk$ derivative inherited from the graphene surface mode, it is now able to cross the branch associated to ZnS, producing the region of highly-efficient modes evident in the plot. According to this first analysis, we would be tempted to state that a lower value of the chemical potential, leading to a lower $d\omega/dk$ derivative of the dispersion relation, would produce an even larger flux, since it would produce the discussed coupling at even larger values of $k$, contributing more flux. Nevertheless, this statement ignores the fact that the modes effectively participating to the radiative heat transfer strongly depend on the distance $d$ through the exponential factor $e^{-2\Ima(k_z)d}$ in the evanescent region [see second line of Eq.~\eqref{Twk}]. Thus, the optimal $\mu$ is, within this simplified approach, the one producing the coupling between the two branches at the highest $k$ participating to the energy exchange, roughly scaling as $1/d$. This explains why the optimal chemical potential increases with the distance $d$ (thus reducing the $k$ at which the coupling is produced) as shown in the inset of Fig.~\ref{Amplification}. This analysis also explains why this effect basically exists only in the near field, since only in this regime high values of the wavevector can be explored and exploited.

This view is further confirmed by the analysis of the dispersion relations of the cavity modes, given by the green lines in Figs.~\ref{Landauer} and \ref{Landauer_g}. These are obtained as the poles of the determinant of the scattering matrix of the cavity, coinciding with the zeros of the denominator of the transmission coefficient given in Eq.~\eqref{Twk}. In the four panels of Fig.~\ref{Landauer_g}, we clearly see the two lower branches coming from the strong coupling between the individual modes of GaAs and graphene. We observe that, as $\mu$ increases, so does the derivative $d\omega/dk$ of the one at higher frequency. This branch crosses the one describing the surface mode of ZnS at the optimal wavevector for $\mu=0.05\,$eV. For higher values of the chemical potential [see Fig.~\ref{Landauer_g}(c) and (d)], we observe the appearance of a further strong coupling between the graphene-GaAs mode and the one of ZnS, with an increased mode participation taking place at smaller $k$, thus producing a smaller radiative flux, as discussed above.

It is interesting to see the effect of a varying chemical potential on the spectral flux $\varphi(\omega)$, defined by the relation
\begin{equation}
 \varphi = \int_{0}^{+\infty}\!\!\!\!\!\!d\omega\,\varphi(\omega).
\end{equation}
In Fig.~\ref{Grid_20nm}(a) the spectral fluxes corresponding to the three dielectric--dielectric configurations are shown. The quasi-monochromatic flux typical of near-field transfer between equal materials is manifest for GaAs--GaAs and ZnS--ZnS, while the mixed configuration Zns--GaAs shows a much broader and lower spectral flux. Figure \ref{Grid_20nm}(b) shows that the lowest values of $\mu$ allow to tailor the spectral flux by creating a peak around the ZnS resonance frequency which considerably approaches the one of the ZnS--ZnS scenario. Besides, the spectral flux is broader in this case, as a result of the $\sqrt{k}$ behavior coming from the presence of graphene. Finally, the highest values of $\mu$ give, as shown in Fig.~\ref{Grid_20nm}(c), an even broader spectral flux, at the expenses of a reduced peak flux at the resonance frequency.

\section{Robustness with respect to\\frequency mismatch}\label{SecRobustness}

Our analysis has revealed so far that graphene is able not only to modulate the radiative heat flux between the two semi-infinite substrates, but also to fully compensate for the frequency mismatch between the two surface resonances. Nevertheless, this study has been performed for a specific choice of the two dielectrics, corresponding to a frequency mismatch $\omega_1-\omega_2\simeq0.1\times10^{14}\,$rad/s. The aim of this section is to study how the tuning and amplification highlighted so far are robust with respect to the frequency mismatch between the two dielectrics. We expect the presence of graphene to have a negative effect in the case of identical dielectrics (and thus of perfect match of surface-resonance frequency), but it is not evident to guess how the possible amplification depends on the mismatch. To this aim, we perform a parametric study in which we artificially modify the Drude-Lorentz model given in Eq.~\eqref{Drude} describing the optical properties of ZnS, by adding a frequency shift $\Delta\omega$ to both $\omega_\text{L}$ and $\omega_\text{T}$. While representing a theoretical study (since a given $\Delta\omega$ does not necessarily represent a real material and the dissipation rate is kept constant), this analysis gives anyway an indication of the existence and extent of the effect considered here as a function of the frequency mismatch. Based on our definition of $\Delta\omega$, the material corresponding to $\Delta\omega=-0.1\times10^{14}\,$rad/s implies a surface resonance matching the one of GaAs.

\begin{figure}[t!]
\includegraphics[width=0.48\textwidth]{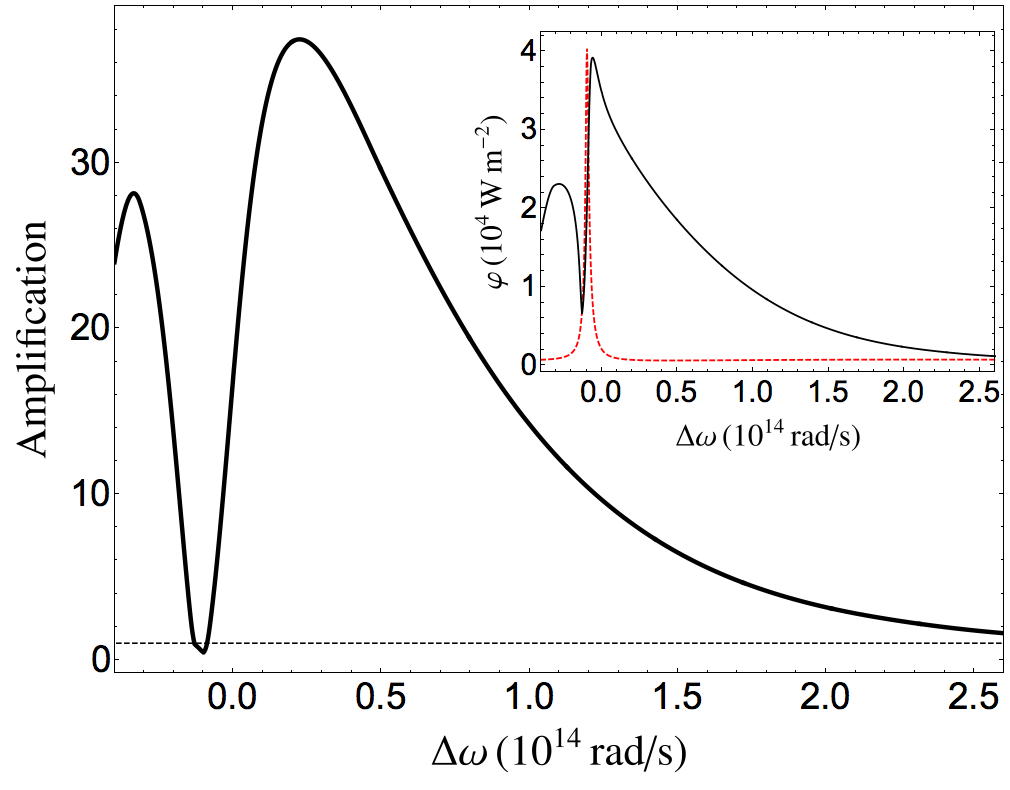} 
\caption{Ratio between the highest flux in the presence of graphene and the one in the absence of graphene as a function of a frequency shift $\Delta\omega$ imposed to the parameters $\omega_\text{L}$ and $\omega_\text{T}$ of ZnS. The horizontal dashed line corresponds to amplification equal to 1. In the inset, the flux in the absence of graphene (red dashed line) and the one in the presence of graphene (black solid line) are shown as a function of $\Delta\omega$.}
\label{Robustness}
\end{figure}

In our analysis we fix the distance to the value $d=20\,$nm and study the amplification (defined as the ratio between the best possible flux in the presence of graphene and the one in the absence of graphene, the one corresponding to the black solid line in Fig.~\ref{Amplification}) as a function of the frequency shift $\Delta\omega$. The results are shown in Fig.~\ref{Robustness}. We first observe that for $\Delta\omega=0$ we find the amplification factor close to 17, already shown in Fig.~\ref{Amplification} for $d=20\,$nm. The transition from $\Delta\omega=0$ to $\Delta\omega=-0.1\times10^{14}\,$rad/s show, as expected, a dramatic reduction of the amplification factor. More specifically, when the frequency shift produces a match between the two individual resonance frequencies, the amplification goes below 1, showing that the presence of graphene is only able to reduce the heat flux in this case. More interestingly, Fig.~\ref{Robustness} shows that, apart from a narrow range of $\Delta\omega$, the amplification shows a high value even for relatively high values of frequency mismatch, proving that this tuning and amplification effect is robust with respect to the choice of materials. This robustness can be understood more in detail by analyzing the inset of Fig.~\ref{Robustness}, where the flux in the absence of graphene (red dashed line) and the optimized one in the presence of graphene (black solid line), i.e. the ones whose ratio gives the amplification factor, are shown. The flux in the absence of graphene shows a very narrow peak as a function of $\Delta\omega$: this confirms that the well-known near-field amplification in the presence of two surface resonances is extremely sensitive to their matching. On the contrary, for a given mismatch $\Delta\omega$, there exists a chemical potential optimizing the flux: as a result, the optimized flux in the presence of graphene remains comparable to the one corresponding $\Delta\omega=0$ for a large range of frequency shifts. Because of the limitation imposed on the values of the chemical potential, the amplification tends to 1 for high values of $\Delta\omega$.

\section{Conclusions}\label{SecConclusions}

We have demonstrated that a graphene sheet can be used as a relay between two dissimilar polar materials which interact in the near field in order to tune and to magnify the radiative heat flux they exchange through the surface phonon-polariton tunneling. This effect results from a coupling of the surface plasmon of graphene with the surface polaritons characterizing the two dielectrics. A direct consequence of this coupling is an increase of the number of modes which contribute to the net flux exchanged between the two materials. More specifically, we have shown that in the near field an optimized choice of the chemical potential is able to produce a flux amplification going beyond one order of magnitude. After discussing this effect in the specific case of GaAs and ZnS, we show that this amplification is robust with respect to the frequency mismatch between surface resonances. Our results broaden further the interest of using graphene in dielectric--dielectric scenarios in order to actively tune radiative heat transfer.


\begin{thebibliography}{99}
%Blackbody theory
\bibitem{Planck} M. Planck, \emph{The theory of heat radiation}, (Dover Publications, New York, 2011).
% Theory of radiative heat transfer
\bibitem{Rytov}S. Rytov, Y. Kravtsov, V. Tatarskii, \emph{Principles of Statistical Radiophysics}, Vol.~3  (Springer-Verlag, Berlin, 1989).
\bibitem{PoldervH}D. Polder and M. van Hove, Phys. Rev. B \textbf{4}, 3303 (1971).
%LDOS above a surface
\bibitem{Eckardt}W. Eckhardt, Z. Physik B\,-\,Condensed Matter \textbf{46}, 85 (1982).
% Theory of radiative heat transfer
\bibitem{JoulainSurfSciRep05}K. Joulain, J.-P. Mulet, F. Marquier, R. Carminati, and J.-J. Greffet, Surf. Sci. Rep. \textbf{57}, 59 (2005).
\bibitem{Volokitin1} A.~I. Volokitin and B.~N.~J. Persson, Rev. Mod. Phys. \textbf{79}, 1291 (2007).
% Experiments on radiative heat transfer
\bibitem{HargreavesPLA69}C. Hargreaves, Phys. Lett. A \textbf{30}, 491 (1969).
\bibitem{KittelPRL05}A. Kittel, W. M\"{u}ller-Hirsch, J. Parisi, S.-A. Biehs, D. Reddig, and M. Holthaus, Phys. Rev. Lett. \textbf{95}, 224301 (2005).
\bibitem{NarayanaswamyPRB08}A. Narayanaswamy, S. Shen, and G. Chen, Phys. Rev. B \textbf{78}, 115303 (2008).
\bibitem{HuApplPhysLett08}L. Hu, A. Narayanaswamy, X. Chen, and G. Chen, Appl. Phys. Lett. \textbf{92}, 133106 (2008).
\bibitem{ShenNanoLetters09}S. Shen, A. Narayanaswamy, and G. Chen, Nano Letters \textbf{9}, 2909 (2009).
\bibitem{RousseauNaturePhoton09}E. Rousseau, A. Siria, G. Joudran, S. Volz, F. Comin, J. Chevrier, and J.-J. Greffet, Nature Photon. \textbf{3}, 514 (2009).
\bibitem{OttensPRL11}R.~S. Ottens, V. Quetschke, S. Wise, A.~A. Alemi, R. Lundock, G. Mueller, D.~H. Reitze, D.~B. Tanner, and B.~F. Whiting, Phys. Rev. Lett. \textbf{107}, 014301 (2011).
\bibitem{KralikRevSciInstrum11}T. Kralik, P. Hanzelka, V. Musilova, A. Srnka, and M. Zobac, Rev. Sci. Instrum. \textbf{82}, 055106 (2011).
\bibitem{KralikPRL12}T. Kralik, P. Hanzelka, M. Zobac, V. Musilova, T. Fort, and M. Horak, Phys. Rev. Lett. \textbf{109}, 224302 (2012).
\bibitem{vanZwolPRL12a}P.~J. van Zwol, L. Ranno, and J. Chevrier, Phys. Rev. Lett. \textbf{108}, 234301 (2012).
\bibitem{vanZwolPRL12b}P.~J. van Zwol, S. Thiele, C. Berger, W. A. de Heer, and J. Chevrier, Phys. Rev. Lett. \textbf{109}, 264301 (2012).
\bibitem{SongNatureNano15}B. Song, Y. Ganjeh, S. Sadat, D. Thompson, A. Fiorino, V. Fern\'{a}ndez-Hurtado, J. Feist, F.~J. Garcia-Vidal, J.~C. Cuevas, P. Reddy, and E. Meyhofer, Nature Nanotechnology \textbf{10}, 253 (2015).
\bibitem{KimNature15}K. Kim, B. Song, V. Fern\'{a}ndez-Hurtado, W. Lee, W. Jeong, L. Cui, D. Thompson, J. Feist, M.~T. Homer Reid, F.~J. Garcia-Vidal, J.~C. Cuevas, E. Meyhofer, and P. Reddy, Nature \textbf{528}, 387 (2015).
\bibitem{StGelaisNatureNano16}R. St-Gelais, L. Zhu, S. Fan, and M. Lipson, Nature Nanotechnology \textbf{11}, 515 (2016).
\bibitem{KloppstecharXiv}K. Kloppstech, N. K\"{o}nne, S.-A. Biehs, A. W. Rodriguez, L. Worbes, D. Hellmann, and A. Kittel, preprint arXiv:1510.06311 (2015).
\bibitem{WatjenAPL16}J.~I. Watjen, B. Zhao, and Z.~M. Zhang, Appl. Phys. Lett. \textbf{109}, 203112 (2016).
%Superplanckian transfer with hyperbolic modes
\bibitem{BenAbdallahPRL12} S.-A. Biehs, M. Tschikin, and P. Ben-Abdallah Phys. Rev. Lett. \textbf{109}, 104301 (2012).
%Transfer with surface Bloch waves
\bibitem{BenAbdallahAPL10} P. Ben-Abdallah, K.Joulain, and A. Pryamikov, Appl. Phys. Lett. \textbf{96}, 143117 (2010).
% Fundamental limits vs Landauer
\bibitem{BenAbdallahPRB10} P. Ben-Abdallah and K. Joulain, Phys. Rev. B \textbf{82}, 121419(R) (2010).
% Graphene and radiative heat transfer
\bibitem{Persson}B.~N.~J. Persson, and H. Ueba, J. Phys. Condens. Matter \textbf{22}, 462201 (2010).
\bibitem{Volokitin}A.~I. Volokitin and B.~N.~J. Persson, Phys. Rev. B \textbf{83}, 241407(R) (2011).
\bibitem{Svetovoy1}V.~B. Svetovoy, P.~J. van Zwol, and J. Chevrier, Phys. Rev. B \textbf{85}, 155418 (2012).
\bibitem{Ilic1}O. Ilic, M. Jablan, J.~D. Joannopoulos, I. Celanovic, H. Buljan, and M. Solja\v{c}i\'{c}, Phys. Rev. B \textbf{85}, 155422 (2012).
\bibitem{Ilic2}O. Ilic, M. Jablan, J.~D. Joannopoulos, I. Celanovic, H. Buljan, and M. Solja\v{c}i\'{c}, Opt. Express \textbf{20}, A366 (2012)
\bibitem{Messina1}R. Messina, J. P. Hugonin, J.-J. Greffet, F. Marquier, Y. De Wilde, A. Belarouci, L. Frechette, Y. Cordier, and P. Ben-Abdallah, Phys. Rev. B \textbf{87}, 085421 (2013). 
\bibitem{Messina2}R. Messina and P. Ben-Abdallah, Sci. Rep. \textbf{3}, 1383 (2013).
\bibitem{Lim2}M. Lim, S.~S. Lee, and B.~J. Lee, Opt. Express \textbf{21}, 22173 (2013).
\bibitem{Phan}A.~D. Phan, S. Shen, and L.~M. Woods, J. Phys. Chem. Lett. \textbf{4}, 4196 (2013).
\bibitem{Liu3}X.~L. Liu and Z. Zhang, Appl. Phys. Lett. \textbf{104}, 251911 (2014).
\bibitem{Liu1}X. Liu, R.~Z. Zhang, and Z. Zhang, ACS Photonics \textbf{1}, 785 (2014).
\bibitem{Svetovoy2}V.~B. Svetovoy and G. Palasantzas, Phys. Rev. Appl. \textbf{2}, 034006 (2014).
\bibitem{Drosdoff}D. Drosdoff, A.~D. Phan, and L.~M. Woods, Advanced Optical Materials \textbf{2}, 1038 (2014).
\bibitem{Zhang1}R.~Z. Zhang, X. Liu, and Z.~M. Zhang, AIP Advances \textbf{5}, 053501 (2015).
\bibitem{Lim1}M. Lim, S. Jin, S.~S. Lee, and B.~J. Lee, Opt. Express \textbf{23}, A240 (2015).
\bibitem{Chang}J.-Y. Chang, Y. Yang, and L. Wang, J. Quant. Spectrosc. Radiat. Transf. \textbf{184}, 58 (2016).
\bibitem{Song}J. Song and Q. Cheng, Phys. Rev. B \textbf{94}, 125419 (2016).
\bibitem{Yin}G. Yin, J. Yang, and Y. Ma, Appl. Phys. Express \textbf{9}, 122001 (2016).
\bibitem{Zheng}Z. Zheng, X. Liu, A. Wang, and Y. Xuan, Int. J. Heat Mass Transfer \textbf{109}, 63 (2017).
\bibitem{Zhao1}B. Zhao and Z.~M. Zhang, ASME J. Heat Transfer \textbf{139}, 022701 (2017).
\bibitem{Simchi}H. Simchi, J. Appl. Phys. \textbf{121}, 094301 (2017).
\bibitem{Lim3}M. Lim, S.~S. Lee, and B.~J. Lee, J. Quant. Spectrosc. Radiat. Transf. (2017, in press).
\bibitem{Zhao2}Q. Zhao, T. Zhou, T. Wang, W. Liu, J. Liu, T. Yu, Q. Liao, and N. Liu, J. Phys. D: Appl. Phys. \textbf{50}, 145101 (2017).
\bibitem{Shi}K. Shi, F. Bao, and S. He, ACS Photonics \textbf{4}, 971 (2017).
% Graphene
\bibitem{Geim1}Geim, A. K. \& Novoselov, K. S. The rise of graphene. \emph{Nat. Mater.} \textbf{6}, 183 (2007).
\bibitem{Geim2}Geim, A. K. Graphene: Status and prospects. \emph{Science} \textbf{324}, 1530 (2009).
%Optical data
\bibitem{Palik}\textit{Handbook of Optical Constants of Solids}, edited by E. Palik (Academic Press, New York, 1998).
% Optical properties of graphene
\bibitem{FalkovskyJPhysConfSer08}Falkovsky, L. A. Optical properties of graphene. \emph{J. Phys. Conf. Ser.} \textbf{129}, 012004 (2008).
\bibitem{JablanPRB09}M. Jablan, H. Buljan, and M. Solja\v{c}i\'{c}, Phys. Rev. B \textbf{80}, 245435 (2009).
\end{thebibliography}
\end{document}